\documentclass[aps,prl,twocolumn,showpacs]{revtex4-1}
\usepackage{amssymb}
\usepackage{graphicx}
\usepackage{amsmath}

\usepackage{times}
\usepackage{color}
\usepackage{subfigure}
\usepackage{setspace}
\usepackage{bm}
\usepackage{ulem}

\begin{document}

\newcommand {\beq} {\begin{equation}}
\newcommand {\eeq} {\end{equation}}
\newcommand {\bqa} {\begin{eqnarray}}
\newcommand {\eqa} {\end{eqnarray}}
\newcommand {\ca} {\ensuremath{c^\dagger}}
\newcommand {\ba} {\ensuremath{b^\dagger}}
\newcommand {\Ma} {\ensuremath{M^\dagger}}
\newcommand {\psia} {\ensuremath{\psi^\dagger}}
\newcommand {\fbar} {\ensuremath{\bar{f}}}
\newcommand {\psita} {\ensuremath{\tilde{\psi}^\dagger}}
\newcommand{\lp} {\ensuremath{{\lambda '}}}
\newcommand{\A} {\ensuremath{{\bf A}}}
\newcommand{\Q} {\ensuremath{{\bf Q}}}
\newcommand{\kk} {\ensuremath{{\bf k}}}
\newcommand{\qq} {\ensuremath{{\bf q}}}
\newcommand{\kp} {\ensuremath{{\bf k'}}}
\newcommand{\rr} {\ensuremath{{\bf r}}}
\newcommand{\rp} {\ensuremath{{\bf r'}}}
\newcommand {\ep} {\ensuremath{\epsilon}}
\newcommand{\nbr} {\ensuremath{\langle r r' \rangle}}
\newcommand {\no} {\nonumber}
\newcommand{\up} {\ensuremath{\uparrow}}
\newcommand{\dn} {\ensuremath{\downarrow}}
\newcommand{\rcol} {\textcolor{red}}

\begin{abstract}
We make the first testable predictions for the
{\it local two-particle} spectral function of a disordered s-wave superconductor, probed by scanning Josephson spectroscopy (sjs),
providing complementary information to scanning tunneling spectroscopy (sts). 
We show that sjs provides a direct map of the local
superconducting order parameter that is found to be anticorrelated with the gap
map obtained by sts. Furthermore, this anticorrelation increases with disorder.
For the momentum resolved spectral function, we find the Higgs mode shows a non-dispersive subgap feature at low momenta, spectrally separated from phase modes, for all disorder strengths. The amplitude-phase mixing remains small at low momenta even when disorder is large. Remarkably, even for large disorder and high momenta, the amplitude-phase mixing oscillates rapidly in frequency and hence do not affect significantly the purity of the Higgs and phase dominated response functions. 

 \end{abstract}
\title{ Two-particle spectral function for disordered s-wave
  superconductors: \\
  local maps and collective modes}
\author{Abhisek Samanta$^1$, Amulya Ratnakar$^2$, Nandini Trivedi$^3$ and Rajdeep 
Sensarma$^1$}
 \affiliation{1. Department of Theoretical Physics, Tata Institute of Fundamental
 Research, Mumbai 400005, India. \\
2. UM-DAE Centre for Excellence in Basic Sciences, Mumbai University,
Mumbai, India\\
3. Physics Department, The Ohio State University, Columbus, Ohio, USA 43201 }

\date{\today}

\maketitle

\noindent{\it Introduction:}
Superconductivity,
characterized by a
  macroscopic complex wavefunction of Cooper pairs, can be
destroyed along two distinct routes: (a) by reducing the amplitude of
the wavefunction to zero, as observed
in conventional clean superconductors at $T_c$, where Cooper pairs break apart,
or (b) by disordering the phase of the wavefunction, while keeping
the pairing
amplitude finite, as seen in strongly interacting ~\cite{Lee_Rev,Plain-Vanilla,BCS-BEC_Mohit},
or in strongly disordered superconductors~\cite{Fisher,
  Nandini1,Nandini2,Nandini3}.

There is strong experimental~\cite{Pratap1,Pratap2,HiggsExpt} and
theoretical ~\cite{Nandini1, Nandini2, Benfatto1,Benfatto2,Nandini3}
evidence that the destruction of superconductivity
in thin films at high disorder ~\cite{Expt1,Expt2,Expt3} is driven by
loss of phase coherence of Cooper pairs, whereas the single particle fermionic spectrum remains gapped through the superconductor to insulator transition. The low
energy excitations of this system are the
dynamical fluctuations of the amplitude (Higgs) and phase (Goldstone) modes of the
complex order parameter. The Higgs mode in superconductors has been
studied experimentally using optical~\cite{Opt_pump} and Raman~\cite{NbSe2_Higgs}
spectroscopy. It has also been studied in neutral ultracold atomic systems through lattice modulation spectroscopy~\cite{Bloch_Higgs}.
In disordered superconductors, recent experiments~\cite{HiggsExpt} have
interpreted low energy optical absorption as indicative of absorption by Higgs modes within the single particle gap.


While the claim of observing pure amplitude Higgs modes in ultracold atoms is undeniable, there are two main issues that prevent current experiments on quantum materials from
reaching similar unambiguous conclusions: (a) Materials are inherently disordered so it is not evident to what extent the low energy
absorption can be separated into pure phase and amplitude (Higgs) modes in systems with broken
translational symmetry. This is one of the key questions we address and answer in this article.
 (b) The experiments currently do not
have direct access to a spatial map of the inhomogeneous superconducting order parameter in the disordered systems.
A systematic
study of the evolution of collective modes with disorder is required
to resolve these issues.

\noindent {\it Main results:}
In this work, we use a non-perturbative functional integral approach~\cite{Diener} to trace the evolution of the
two-particle collective spectrum of a disordered attractive Fermi
Hubbard model. 
We present for the
first time the full momentum and frequency dependence of the disorder
averaged spectral function as well as spectral function maps in real
space for a given disorder realization.
Our spectral function maps at large disorder show strong correlation between
superconducting patches and low energy pair spectral weight, which are found to be anticorrelated with regions of
large local gap. We thus make testable predictions for scanning tunneling~\cite{Sacepe} and scanning Josephson spectroscopies~\cite{Mallika} for the first time.

From our theoretical approach we can easily separate the
contribution of the amplitude (Higgs) modes, the phase modes and the
amplitude-phase mixing. We therefore obtain key insights into the relative importance of the mixing contributions  in different spectral regions and for different degrees of disorder.
We find that the local response is dominated by the phase modes, while the Higgs and amplitude-phase mixing contributions play a subdominant role. An intriguing feature of the spatial maps of the amplitude-phase mixing contribution is its oscillatory nature over length scales much shorter than the superconducting patches. Since the tunneling probes average over a few lattice spacings, we expect the mixing to not be important.

The momentum-dependent spectral functions show two features important for understanding the experiments: (a) At arbitrarily low disorder, the Higgs modes at low momenta form non-dispersive states below the two particle threshold
that are spectrally separated from the low energy phase modes. This subgap feature persists with increasing disorder and is predominantly made of Higgs fluctuations with small amplitude-phase mixing. Thus, we expect experiments observing subgap features at finite frequency are indeed probing the Higgs modes of the disordered system. (b) The amplitude phase mixing at higher momenta show a dramatic evolution with disorder. At low disorder it is predominantly of one sign, while at larger disorder, it oscillates and changes sign rapidly as a function of frequency. Thus, we expect that high disorder, temperature, or finite resolution broadening of spectroscopic probes, will washout the effects of amplitude-phase mixing, a result that is rather counter-intuitive.
In summary, our work makes testable predictions for
experiments and provides a bridge between microscopic models that start
with fermionic degrees of freedom~\cite{Nandini1,Nandini2} and
effective bosonic models~\cite{Fisher, Sachdev}.

\medskip

\noindent {\it  Model and Methods:}
We analyze the behavior of the disordered attractive Hubbard model on a square
lattice using an inhomogeneous self-consistent functional integral approach. The Hamiltonian is given by:
\beq
\label{eq:Ham}
H=-t\sum_{\nbr\sigma}\ca_{r\sigma}c_{r' \sigma}-U\sum_rn_{r\up}n_{r\dn}+\sum_r v_rn_r
\eeq
where $c^{\dagger}_{r\sigma}(c_{r\sigma})$ is the creation (annhilation) operator for 
electrons with spin 
$\sigma$ on site $r$, $t$ is the nearest-neighbour hopping, and $|U|$ is the attractive interaction leading to Cooper pairing. 
Here, $v_r$ is a random potential, drawn
independently for each site from a uniform distribution of  zero mean and
width $V$, where $V$ sets the scale of disorder in the problem. 

Within our approach, the mean field theory is the saddle point of the
fermionic action, described by static local Cooper pairing field $\Delta_0(r) =\langle
c^\dagger_{r\up}c^\dagger_{r\dn}\rangle$ and the Hartree shift
$\xi_0(r)=U\langle c^\dagger_{r\sigma}c_{r\sigma}\rangle$, determined self-consistently.
We expand the action around this saddle point by considering a fluctuating pairing field
$\Delta(r,\tau)=(\Delta_0(r) +\eta(r,\tau))e^{i\theta(r,\tau)}$ upto
quadratic order, where
$\eta$ and $\theta$ correspond to the amplitude and phase fluctuations
of the order parameter, and obtain,
\bqa
\label{eq:P}
\nonumber {\cal P}_{11}(r,r',\omega)&=&
-\frac{1}{\pi} \text{Im}\langle
\eta(r,\omega+i0^+)\eta(r',-\omega+i0^+)\rangle \\
{\cal P}_{12}(r,r',\omega)&=&
-\frac{1}{\pi} \text{Im}\langle
\eta(r,\omega+i0^+)\theta(r',-\omega+i0^+)\rangle\\ 
\nonumber {\cal P}_{22}(r,r',\omega)&=&
-\frac{1}{\pi} \text{Im}\langle
\theta(r,\omega+i0^+)\theta(r',-\omega+i0^+)\rangle
\eqa
where ${\cal P}_{11}$ is the spectral density of amplitude fluctuations,
${\cal P}_{22}$ that of phase fluctuations and ${\cal P}_{12}$ is the
amplitude-phase mixing term~\cite{analytic_cont} [See SM for details]. 

While $\eta$ and $\theta$ are the natural choice of
fluctuation co-ordinates,
experimental probes, which couple to the fermion density or current, always couple
to $\Delta_0(r)
e^{i\theta(r,\tau)}\sim i \Delta_0(r) \theta(r,\tau)$. The two
particle correlation function, measureable by Josephson spectroscopy,
is $P(r,r',\omega)=\sum_{\alpha\beta}
P_{\alpha\beta}(r,r',\omega)$, where $P_{11}={\cal P}_{11}$,
$P_{12}(r,r',\omega)=\Delta_0(r){\cal P}_{12}(r,r',\omega)$, $P_{21}(r,r',\omega)=\Delta_0(r'){\cal P}_{21}(r,r',\omega)$ and
$P_{22}(r,r'\omega)=\Delta_0(r)\Delta_0(r'){\cal
  P}_{22}(r,r',\omega)$. We will now consider the evolution of these experimentally
measurable spectral functions with disorder.

\medskip

\noindent {\it Local pair spectral function:}
In a mean field description, the
system breaks up into superconducting and insulating
islands at intermediate and large disorder~\cite{Nandini1}. STM
measurements also show indirect evidence of strong spatial
inhomogeneity in patchy single particle gap-maps~\cite{Sacepe}. However, a direct access to the
inhomogeneous superconducting order parameter is missing in these
systems. We find that the integrated local two-particle spectral weight is strongly spatially correlated with the superconducting
order parameter and further shows strong anti-correlation with
local single particle gaps. Our prediction can be experimentally tested by combining
scanning tunneling with scanning Josephson spectroscopy data~\cite{Mallika}.



In Fig.~\ref{localmap}(a) and (b), we show the local order parameter
$\Delta_0(r)$ and the integrated
local 2-particle spectral weight $F(r) =\int_0^{2E_{gap}}
P(r,r,\omega)$ for a typical configuration at
large disorder ($V=6$)~\cite{Egap}. We notice the strong
spatial correlation between regions with large $\Delta_0(r)$ and large
$F(r)$. Although regions with small $\Delta_0(r)$ have
small phase stiffness, these phase fluctuations do not contribute
to the pair spectral function as $\Delta_0(r)$ is small in these
regions. We have checked that this strong correlation is robust to
choice of disorder configurations and to variation of cutoffs used to
calculate $F(r)$ [See SM for details]. The integrated spectral weight can thus be used to
experimentally map out the superconducting regions in the
system. In Fig.~\ref{localmap}(c), we plot the local single particle
gap $E(r)$, obtained from peaks in the  local one particle density of states
 for the same configuration [See SM for details]. The maps in Fig.~\ref{localmap}
(a) and (b) shows strong spatial anti-correlation
between regions with large $\Delta_0(r)$ or $F(r)$ and regions with large
$E(r)$, i.e. large single particle gaps map out the
insulating regions in the system. To track the evolution of this
strong anti-correlation between $F(r)$ and $E(r)$, in
Fig.~\ref{localmap}(d), we plot the
covariance of these quantities, averaged over disorder configurations,
as a function of $V$. The negative correlations
increase with disorder, as the system breaks up into superconducting
and non-superconducting regions.  

The relative contribution of the Higgs mode ($P_{11}$), phase mode $P_{22}$, and the
amplitude-phase mixing ($P_{12}+P_{21}$) to the 2-particle spectral
function is a key question of interest, especially in the light of
papers with contradictory claims on this
matter~\cite{Auerbach,Pratap1,Benfatto1, Pratap3}. 
In
Fig.~\ref{localmap}(e) we plot $F_{11}+F_{22}$ as bars on each lattice
site with the contribution from $F_{22}$ shown in red and that from
$F_{11}$ shown in yellow. We find that the local 2-particle spectral weight is
dominated by the phase modes, with the amplitude and mixing
contributions playing a subleading role. The contribution of the
mixing, $F_{12}+F_{21}$ is plotted as a map in
Fig.~\ref{localmap}(f). $P_{12}$ and $P_{21}$ does not have the
interpretation of a spectral weight and changes from positive to
negative over different regions in the map. Hence, while
mixing plays a somewhat important but subleading role in the local
spectral weight, it should have minimal impact on the signals in
probes which look at spatially averaged quantities. 

\medskip

\noindent {\it Momentum and energy dependence of collective modes:}
We now
consider the spectral function $P_{\alpha\beta}(q,\omega)=\sum_{rr'}
e^{i q \cdot (r-r')} P_{\alpha\beta}(r,r',\omega)$ (after disorder
averaging) to study the behaviour of the collective modes. In the optical conductivity the pair spectral function contributes to loop corrections, hence their effect cannot be spectrally resolved. A more direct momentum and frequency resolved measurement is possible with the recently developed M-EELS techniques~\cite{Abbamonte}. 
Fig.~\ref{Pqw}(a) -(d) shows the Higgs spectral
function $P_{11}(q,\omega)$ with increasing disorder. For $V=0$, a
Goldstone mode exists, but
the Higgs contribution to the spectral weight vanishes as $q\rightarrow
[0,0]$. The picture changes dramatically even for a weak disorder
of $V=0.1t$, where the Higgs mode develops finite weight at the
zone center at an energy well below the two-particle continuum
threshold. We have checked that this phenomenon exists even at a
weaker disorder of $V=0.05$. The relatively flat dispersion of the
Higgs mode suggests localization of these modes at a
finite energy. With increasing disorder the Higgs mode
flattens and broadens, with the threshold for the mode decreasing
with disorder. The Higgs threshold is plotted as a function of
disorder in Fig.~\ref{Pqw}(i). It does not
follow the continuum threshold ($2E_{gap}$) even at low disorder. We also observe a pile up of low energy weight at the
commensurate M point ($[\pi,\pi]$) at intermediate disorder of
$V=1.0$, indicating fluctuating pair-density waves,
although there is no zero energy weight and static
order is absent in the mean-field theory.

We now focus on the phase contribution to the spectral function
$P_{22}(q,\omega)$ in Fig~\ref{Pqw} (e)-(h).
The linearly dispersing collective
mode at low $q$ broadens with disorder, and the dispersion becomes flatter. The dispersive
mode can be identified even for large disorder
$V\approx 5t$. The speed of sound, extracted from the slope of the
dispersion, is plotted in Fig.~\ref{Pqw} (i). It decreases with
disorder, going to zero near $V\approx5.5 t$.
It is also evident from the color-scales that phase fluctuations dominate over
amplitude fluctuations in the entire disorder range.
Finally, in Fig~\ref{Pqw} (j)-(l), we plot the mixing term
$P_{12}(q,\omega)+P_{21}(q,\omega)$, as a function of $q$ and
$\omega$ for increasing disorder. It is evident that with increasing
disorder the mixing term rapidly oscillates between positive and
negative values as a function of the frequency and hence mixing terms
give small contributions to the pair spectral function.

The clear dominance of the phase modes over Higgs modes
leads to the
question whether the interesting features of the Higgs spectral
function can be visible in experiments. Fortunately, the features of
the Higgs and the phase spectral functions are well separated in
energy at low $q$ and hence probes which couple to the
spatially averaged pair spectral function in a energy resolved manner
should see these features clearly (see
Fig~\ref{Pw}(a)-(f)).
It is important to note that this spectral separation is a feature of low $q$ response and is
not present in the local response we investigated in the previous section. 
We also find that the amplitude-phase mixing term has negligible
contribution at all frequencies near $q=0$, and the relative contribution
decreases with disorder, contrary to the popular
belief that they are the dominant force in shaping the collective
spectrum. This can be understood from the fact that the mixing
contribution varies from positive to negative values in space, as seen
in Fig.~\ref{localmap}(f), and hence averages to zero when one looks
at low $q$ response of the system.

{\it Discussion:}
We have investigated the evolution of collective modes in a disordered
s-wave superconductor starting from a microscopic description. We find
that the local 2-particle spectral weight is strongly
correlated with the superconducting regions and strongly
anti-correlated with regions of high one particle spectral gap. The
pair response is dominated by the phase mode, but the Higgs mode shows
interesting features at low $q$ which are spectrally separated from
the phase mode contributions. The amplitude phase mixing term plays a
subdominant role at large disorder due to rapid change of sign.

\begin{acknowledgements}
The authors thank P. Raichaudhuri for useful discussions. A.S.,
A.R. and R.S. acknowledge computational facilities at the Department
of Theoretical Physics, TIFR Mumbai. N.T.
acknowledges funding from grant
NSF-DMR-1309461. 

\end{acknowledgements}


\begin{thebibliography}{99}
\bibitem{analytic_cont} We note that our formalism calculates the
correlators in real frequencies analytically and does not suffer from
issues of analytic continuation.
\bibitem{Egap} $E_{gap}$ is the gap to the single particle Fermionic
  excitations, and $2E_{gap}$ is the threshold for continuum of
 2-particle  excitations. We restrict the integral upto $2E_{gap}$ so that only
contribution of collective modes are considered.
\end{thebibliography}



\begin{figure*}
\centering

\includegraphics[width=0.9\textwidth]{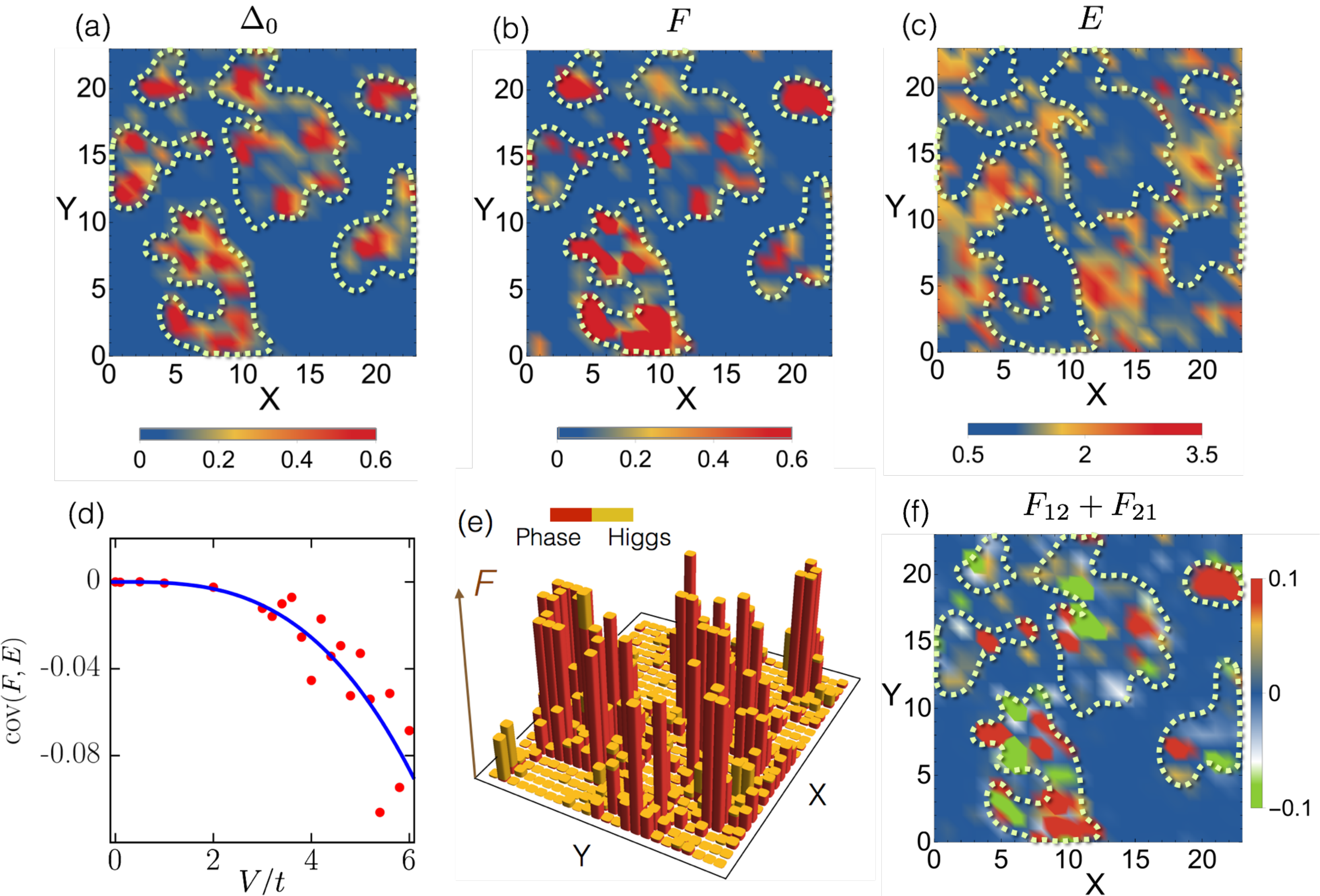}
  \caption{Spatial maps for a particular disorder configuration at
    $V=6t$ showing
  strong correlation between (a) the local superconducting order parameter
  $\Delta_0(r)$ and (b) the frequency-integrated local 2-particle spectral
  weight $F(r)$. (c) Spatial map of the corresponding single particle gap $E(r)$ obtained
  from the local 1-particle density of states [See SM for details]. Notice the strong
  anticorrelation between (c) and (b). (d) Covariance between $F(r)$
  and $E(r)$, averaged over disorder realizations, as a function of disorder
  strength. The anticorrelation increases with disorder. (e): A bar map showing
  the relative weights of the Higgs (amplitude) and Goldstone (phase) modes in the two-particle spectral weight $F(r)$ shown in (b). The
  spectral weight at large disorder is dominated by the phase
  modes. (f) The integrated amplitude-phase mixing two-particle spectral weight $F_{12}(r)+F_{21}(r)$
  corresponding to the configuration shown in (b). The mixing
  contribution shows regions with positive and negative values on a scale much smaller than the superconducting coherence length.}
\label{localmap}
\end{figure*}

\begin{figure*}
\centering
\includegraphics[width=1.0\textwidth]{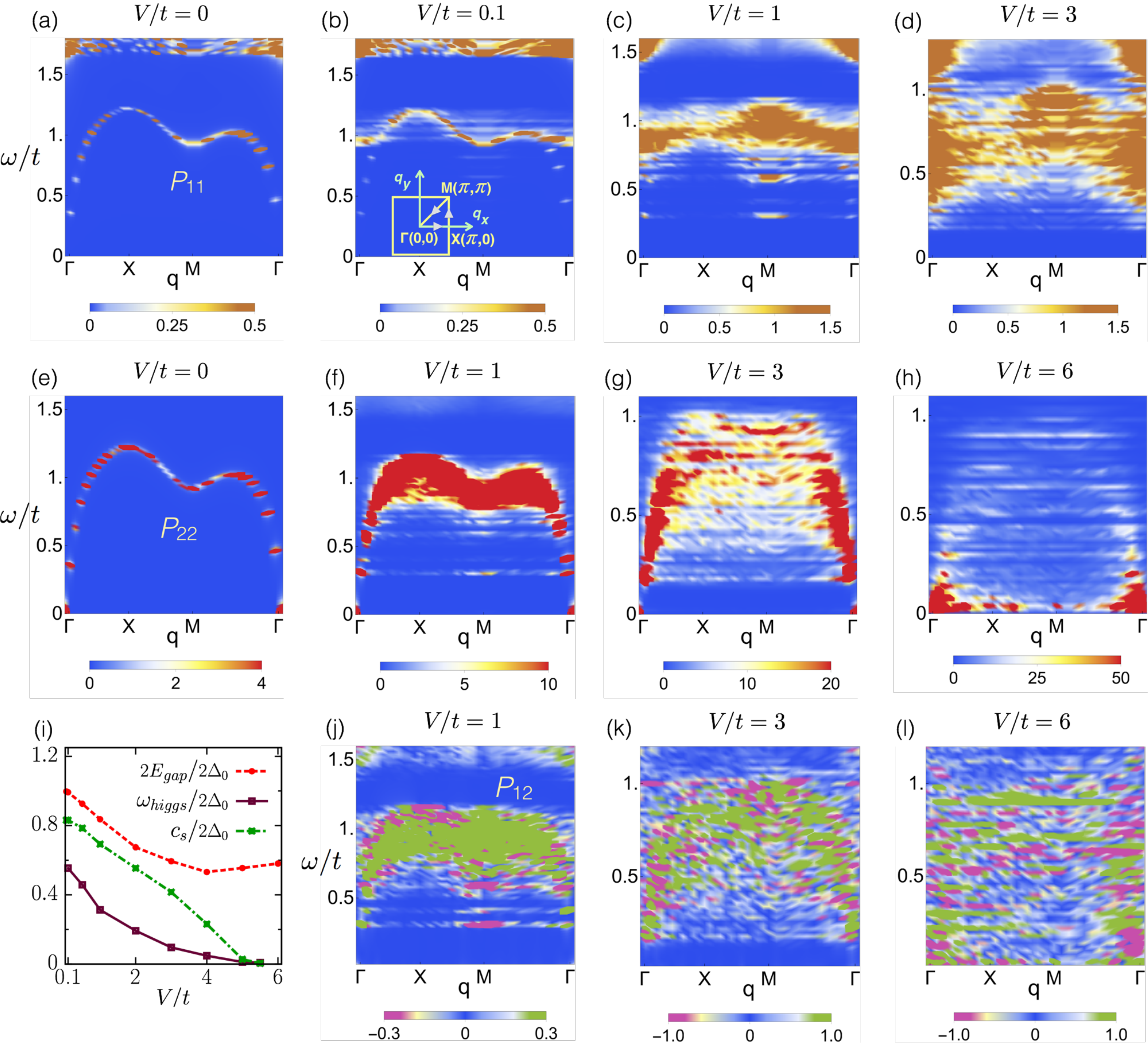}
\caption{ (a)-(d): Higgs spectral function $P_{11}(q,\omega)$ [Eqn.~\ref{eq:P}] shown as a density plot in $q$ and
  $\omega$ with increasing disorder: (a) $V=0.0 t$ showing no weight at $q=0$; (b) weak disorder $V=0.1 t$ starts showing finite
  $q=0$ weight of the Higgs mode; (c) $V=1.0 t$ and (d) $V=3.0 t$.  (e)-(h): Goldstone or phase spectral function $P_{22}(q,\omega)$ [Eqn.~\ref{eq:P}] shown as a density plot in $q$ and 
  $\omega$ for (e) $V=0.0 t$ (f) $V=1.0 t$ (g) $V=3.0 t$ and (h) $V=6.0 t$. Note the
  relative stability of dispersive modes up to large disorder
  strength. (i) The Higgs threshold $\omega_{higgs}$, the speed of
  sound $c_s$ and the two particle continuum threshold $2E_{gap}$ (pair-breaking scale) as a
  function of $V$.
(j)-(l): The amplitude-phase mixed two-particle spectral function
$P_{12}+P_{21}$ [Eqn.~\ref{eq:P}] shown as a density plot as a function of $q$ and $\omega$ for (j) $V=1.0 t$ (k)
$V=3.0 t$ and (l) $V=6.0 t$. Note that the mixing term  grows in magnitude but oscillates in
sign more rapidly as disorder is increased leading to cancellations in measurable response functions. }
\label{Pqw}
\end{figure*}

\begin{figure*}
\centering
\includegraphics[width=0.9\textwidth]{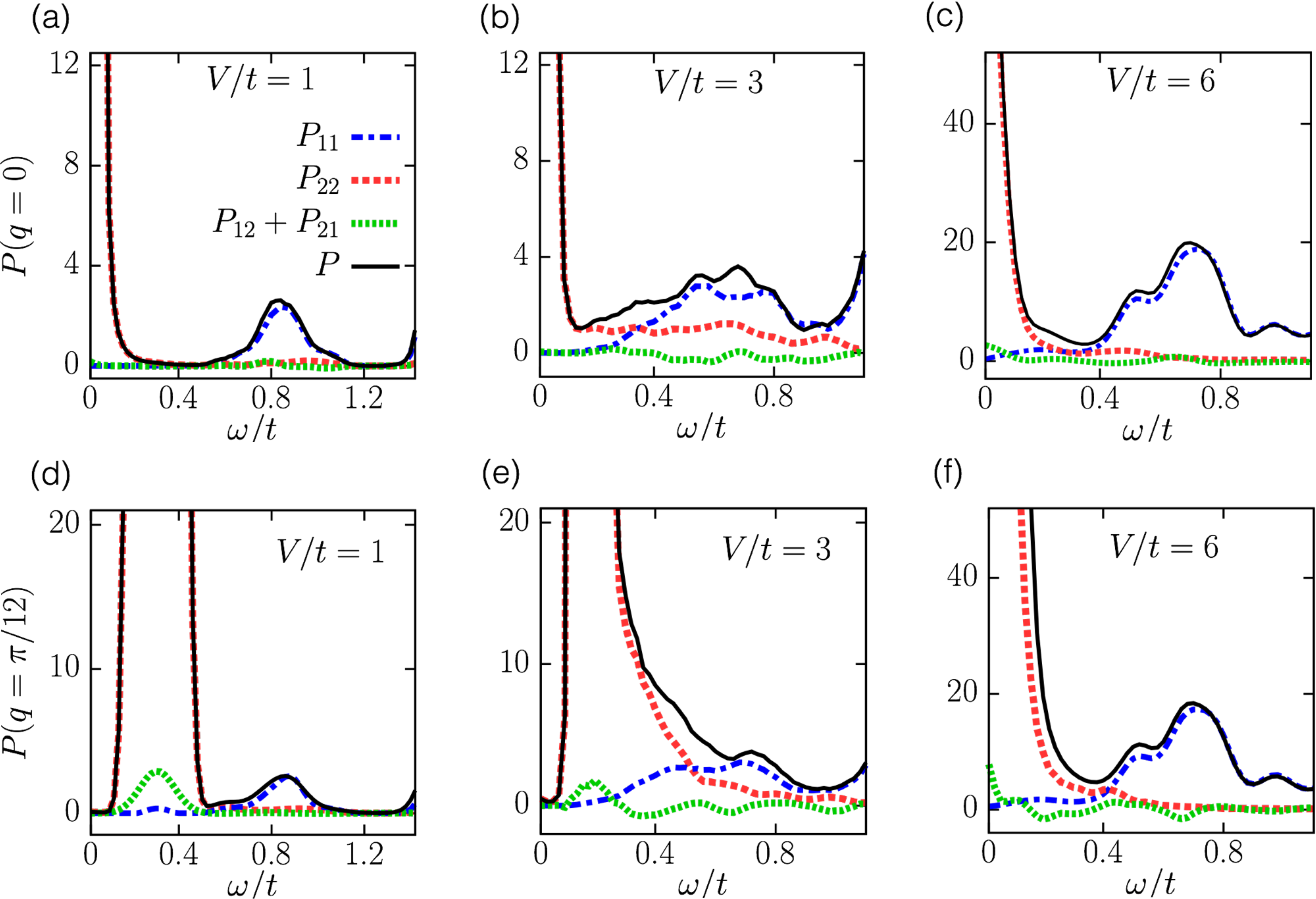}
\caption{ Energy dependence of the spectral function $P$ for increasing disorder $V=1.0 t$, $V=3.0 t$ and $V=6.0 t$ for  (a)-(c): $q=[0,0]$ and (d)-(f): $q=[\pi/12,0]$. The decomposition of the spectral function contributions from the
Higgs mode $P_{11}$, the phase mode $P_{22}$ and the mixing $P_{12}+P_{21}$ is also shown. Note the negligible mixing contributions, and the spectral separation of Higgs and phase contributions. At low disorder the weight of the mixing term is similar to
that of the Higgs mode, although they are spectrally separated. At
larger disorder, the mixing term is weaker than the Higgs weight,
and both are much smaller than the weight in the phase mode.}
\label{Pw}
\end{figure*}


%

\begin{widetext}\clearpage
\begin{center}
\textbf{\large  Supplementary Material for:  Two-particle spectral function for disordered s-wave
  superconductors: \\
  local maps and collective modes}
\end{center}
\end{widetext}

%

\maketitle
\section{Functional Integral Approach}
We will briefly sketch the key steps used to calculate the
Higgs (${\cal P}_{11}(r,r',\omega)$ in the main text) and the phase
spectral function (${\cal P}_{22}(r,r',\omega)$) as well as the
amplitude-phase mixing term (${\cal P}_{12}(r,r',\omega)$) in the
disordered s-wave superconductor within a functional integral
approach. The partition function $Z$ for the disordered negative $U$
Hubbard model (Eqn. (1) in main paper) can be written in terms of the 
 fermion fields ($\bar f_{\sigma}(r,\tau), f_{\sigma}(r,\tau)$) as 
 $Z = \int D[\bar{f}_\sigma,f_\sigma]e^{-S[\bar{f}_\sigma,f_\sigma]}$
, with the imaginary time ($\tau$) action 
\begin{widetext}
\begin{equation}
 S = \int_0^\beta d\tau \sum_{rr',\sigma} \bar{f}_{\sigma}(r,\tau) 
 \left[ \partial_\tau\delta_{rr'}+ H^0_{rr'}\right]f_{\sigma}(r',\tau)
 -U\sum_r\bar{f}_{\up}(r,\tau)\bar{f}_{\dn}(r,\tau)f_{\dn}(r,\tau)f_{\up}(r,\tau)
\end{equation}
\end{widetext}
 where $\beta=1/T$, $T$ being the temperature of the system. 
 Using Hubbard-Stratanovich auxilliary field fields $\Delta(r,\tau)$ coupling to 
 $\bar{f}_\up(r,\tau)\bar{f}_\dn(r,\tau)$ and $\xi(r,\tau)$ coupling to 
 $\bar{f}(r,\tau)f(r,\tau)$,  and introducing the Nambu spinors $\psi^\dagger(r,\tau)=\{\bar{f}_{\up}(r,\tau), f_{\dn}(r,\tau)\}$, we get
$ Z = \int D[\bar{f}_\sigma, f_\sigma] D[\Delta^\ast, \Delta] D[\xi]
 e^{-S_{eff}[\bar{f}_\sigma,f_\sigma,\Delta^\ast,\Delta,\xi ]}$, with

\begin{widetext}
\begin{eqnarray}
\label{eq:action}
 S_{eff} &=& \int_{0}^{\beta} d\tau \sum_r \frac{|\Delta(r,\tau)|^2 + |\xi(r,\tau)|^2}{U} - \int d\tau d\tau^{'}\sum_{rr'} \psi^\dagger(r,\tau)
 {G}^{-1}(r,\tau;r',\tau^{'})\psi(r',\tau^{'}),~~~ \text{and}\\
\nonumber {G}^{-1}(r,\tau;r',\tau^{'})&=& \delta(\tau-\tau^{'})\left(\begin{array}{cc}%
 - (\partial_\tau -\mu(r,\tau) ) \delta_{rr'} + t\delta_{\langle rr' \rangle} 
 & -\Delta(r,\tau)\delta_{rr'} \\
 -\Delta^\ast(r,\tau)\delta_{rr'} & - (\partial_\tau +\mu(r,\tau) ) \delta_{rr'}
 - t\delta_{\langle rr' \rangle}
 \end{array}\right),
 \end{eqnarray}
\end{widetext}
 where $\mu(r,\tau) = \mu-v(r)-\xi(r,\tau)$. The static but spatially
 dependent saddle point profile, 
 $\Delta(r,\tau) = \Delta_0(r)$ and $\xi(r,\tau) = \xi_0(r)$,
 reproduce the BdG mean field theory, with
 the saddle point equations $\delta S/\delta \Delta_0(r)=0$ and 
 $\delta S/\delta \xi_0(r)=0$ giving the BdG self-consistency
 equations,
\begin{gather}
 \Delta_0(r) = |U|\sum_n u_n(r)v_n^*(r), \no\\
 \xi_0(r) = |U|\sum_n |v_n(r)|^2 \quad\text{and} ~~\langle n\rangle = \frac{2}{N_s}\sum_{n,r} |v_n(r)|^2
 \label{eq:selfc}
\end{gather}
 where $[u_n(r),v_n(r)]$ are the eigenvector of $G^{-1}(r,r',\omega)$ corresponding to eigenvalue $\omega-E_n$ and 
 $n$ runs over only positive eigenvalues ($E_n>0$).
\begin{figure*}
 \includegraphics[width=0.28\textwidth]{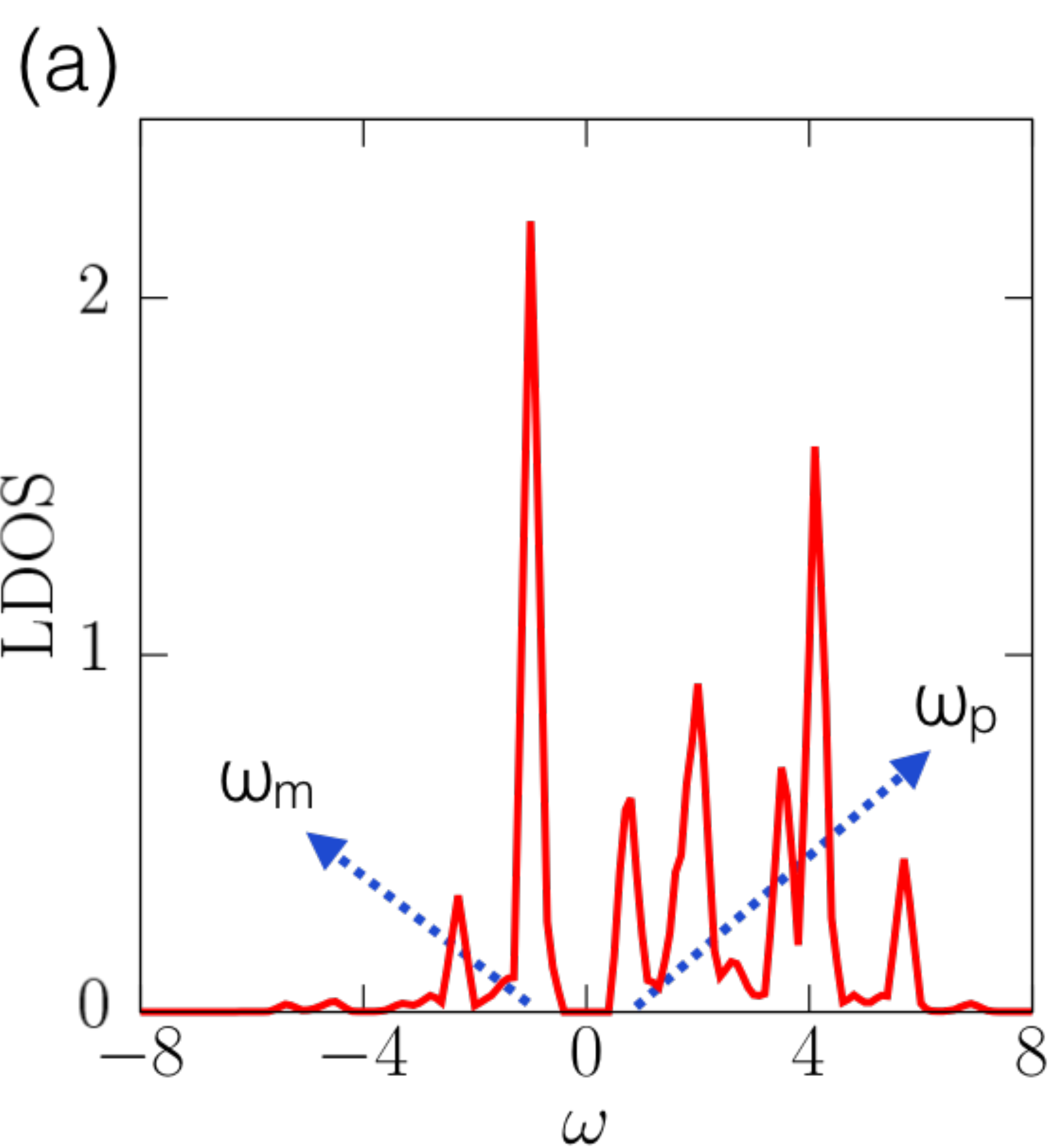}~~~~~~~
 \includegraphics[width=0.305\textwidth]{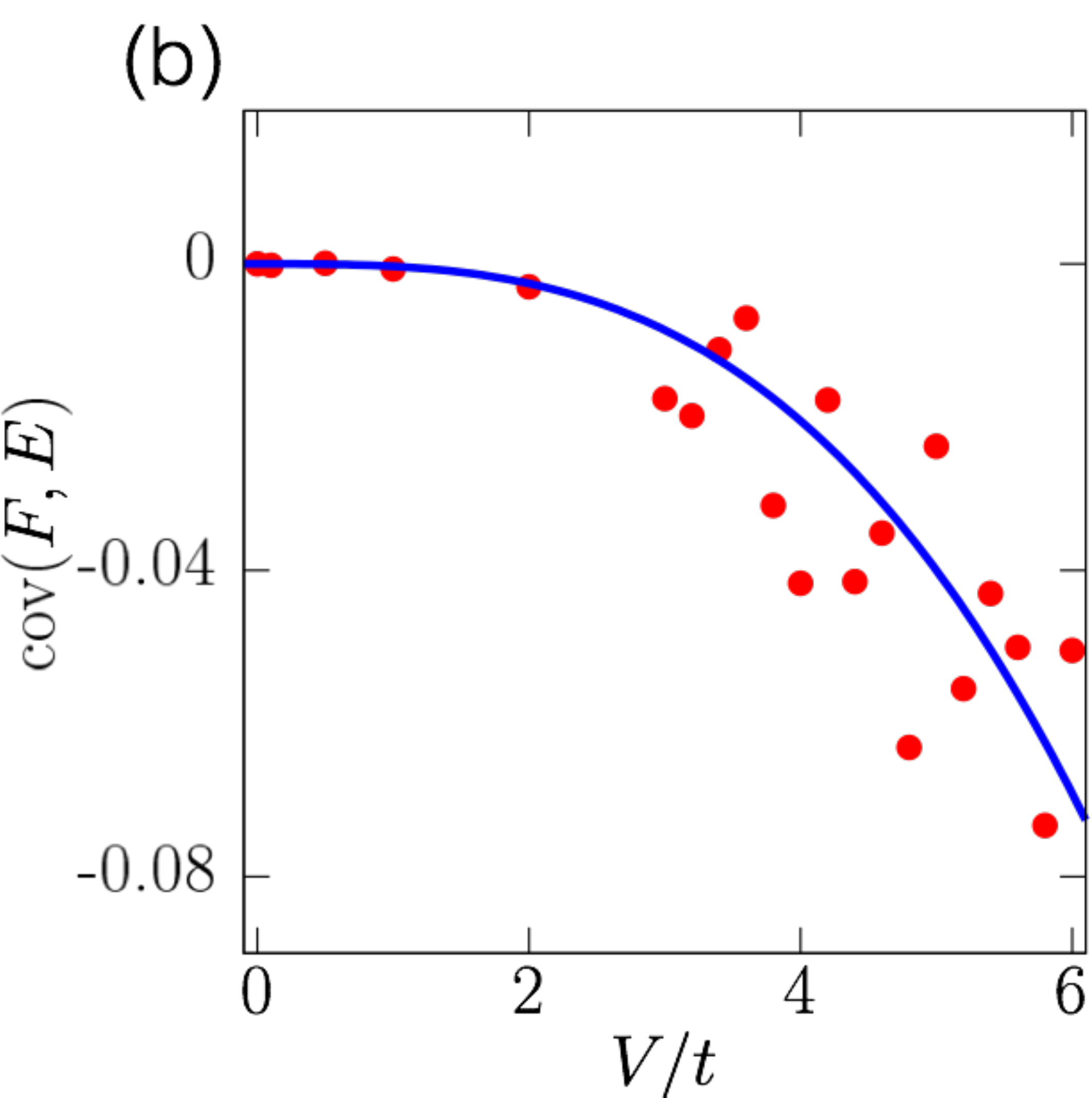}~~~~~~
 \includegraphics[width=0.305\textwidth]{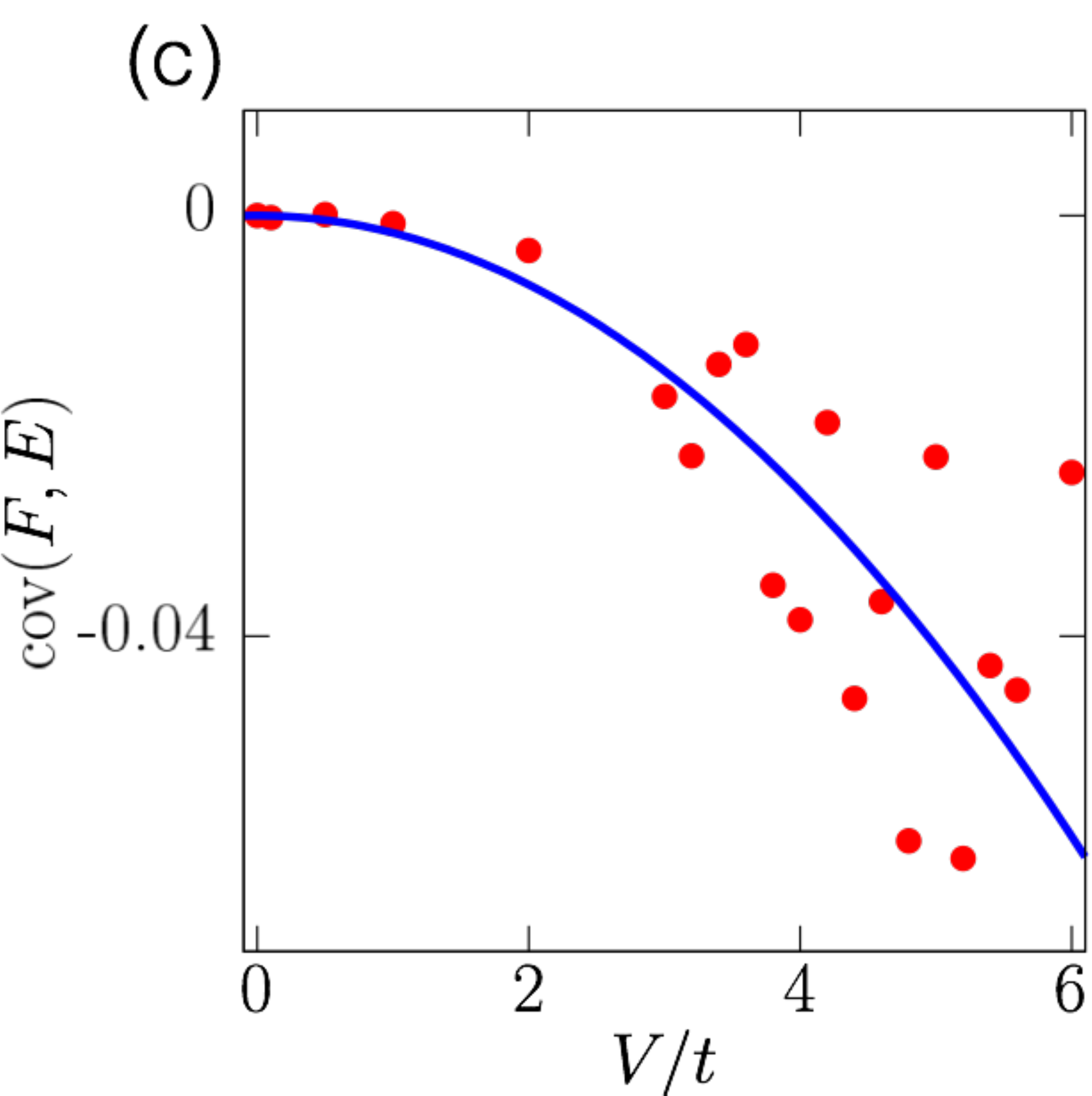}
 \caption{ (a): LDOS as a function of $\omega$ for a particular site. The locations of 
 $\omega_p$ and $\omega_m$ are indicated in the figure. (b)-(c): Covariance between 
 two particle spectral function ($F$) and local single particle gap ($E$) where $F$ is 
 defined as (b) $F(i)=\int_{0.2\times2E_{gap}}^{2E_{gap}} P(i,i,\omega) d\omega$ and 
 (c) $F(i)=\int_{0.3\times2E_{gap}}^{2E_{gap}} P(i,i,\omega) d\omega$. }
 \label{corr}
\end{figure*}

 Going beyond the saddle point approximation, we include (in Eqn. (~\ref{eq:action}) the spatio-temporal fluctuations 
 of the $\Delta$ field i.e.  $\Delta(r,\tau)=(\Delta_0(r) + \eta(r,\tau))e^{i\theta(r,\tau)}$, where $\eta(r,\tau)$ is 
 the amplitude and $\theta(r,\tau)$ is the phase fluctuation around the saddle point
 solution.
 The fermion fields are then integrated out, and the resulting action
 is expanded upto quadratic order in $\eta$ and $\theta$ to obtain the
 Gaussian action for the amplitude and phase fluctuations,
\begin{widetext}  
 \begin{equation}
S_{G} = \sum_{ij}\sum_{\omega_m} \left(\begin{array}{cc}\eta(i,\omega_m)&
 \theta(i,\omega_m)
 \end{array}\right)
 \left(\begin{array}{cc}
 {D^{-1}}_{11}(i,j,\omega_m) &
 {D^{-1}}_{12}(i,j,\omega_m) \\
 {D^{-1}}_{21}(i,j,\omega_m) &
 {D^{-1}}_{22}(i,j,\omega_m) 
 \end{array}\right)
 \left(\begin{array}{cc}
 \eta(j,-\omega_m) \\
 \theta(j,-\omega_m)
 \end{array}\right).
\end{equation}
\end{widetext}
 where $\omega_m=(2m)\pi/\beta$ is the Bosonic Matsubara
 frequency. We work in the amplitude and phase degrees of freedom
 rather than the ``Cartesian'' co-ordinates which mix these degrees of
 freedom, so that we can cleanly talk about Higgs and phase modes. The inverse propagator matrix $D^{-1}$ is analytically
 continued to real frequencies. We note that we work 
 directly in real frequencies and do not need to do numerical analytic continuation. 
 Working at $T=0$, the real frequency retarted inverse 
 fluctuation propagators ${D^{-1}}_{\alpha\beta}(i,j,\omega)$ in terms of BdG eigenvalues 
 and eigenfunctions can be written as:
\begin{eqnarray}
 D^{-1}_{11}(r,r',\omega)
 &=& \frac{1}{U}\delta_{rr'} \\
\nonumber  &+&\frac{1}{2}\sum_{E_{n,n'}
   >0}f^1_{nn'}(r)f^1_{nn'}(r') \chi_{nn'}(\omega)
\end{eqnarray}
where $f^1_{nn'}(r)=\left[u_n(r)u_{n'}(r)-v_n(r)v_{n'}(r)\right]$
and $\chi_{nn'}(\omega)=(\omega+i0^+-E_n-E_{n'})^{-1}
  -(\omega+i0^++E_n+E_{n'})^{-1}$. The off diagonal element is given by
\begin{eqnarray}
 D^{-1}_{12}(r,r',\omega) = -\frac{i\omega}{4} \sum_{E_{n,n'}
   >0} f^1_{nn'}(r)f^2_{nn'}(r') \chi_{nn'}(\omega)~~~~~~
\end{eqnarray}
where $f^2_{nn'}(r)= \left[u_n(r)v_{n'}(r)+v_n(r)u_{n'}(r)\right]$ and
\begin{eqnarray}
 D^{-1}_{22}(r,r',\omega) &=&\tilde{D}_{dia}(r,r')+ 
  \omega^2\kappa(r,r',\omega)
+\Lambda(r,r',\omega)~~~~~~~
 \end{eqnarray}
 where the diamagnetic piece
 $\tilde{D}_{dia}=2\sum_{\langle r r_1\rangle}
   S(r,r_1)$ for $r=r'$, $\tilde{D}_{dia}=-2 S(r,r')$ when $r$ and
   $r'$ are nearest neighbours, and $0$ otherwise,
 where $S(r,r') = \frac{t}{4}\sum_{E_n>0} v_n(r)v_n(r')$, and the
 compressibility 
\begin{eqnarray}
\kappa(r,r',\omega) = \frac{1}{8}\sum_{E_{n,n'} >0}
 f^2_{nn'}(r)f^2_{nn'}(r')\chi_{nn'}(\omega)
\end{eqnarray}
Finally the paramagnetic current-current correlator on the lattice,
$\Lambda(r,r',\omega)$ is given by the expression
\begin{eqnarray}
\nonumber \Lambda(r,r',\omega) &=&\sum_{\langle r r_1\rangle \langle r' r_2\rangle}
J(r,r_1,r',r_2,\omega)-J(r,r_1,r_2,r',\omega)\\
& -&J(r_1,r,r',r_2,\omega)+J(r_1,r,r_2,r',\omega) 
\end{eqnarray}
where
\begin{eqnarray}
 J(r,r_1,r',r_2,\omega) 
 = -\frac{t^2}{8} \sum_{n,n'} f^3_{nn'}(r,r')f^3_{nn'}(r_2,r_1)\chi_{nn'}(\omega) \no\\
\end{eqnarray}
where
$f^3_{nn'}(r,r')=\left[u_n(r)v_{n'}(r')-v_n(r)u_{n'}(r')\right]$.

 We construct the inverse propagators in real space (continued to real frequency),
 invert the matrix to obtain the propagators $D_{\alpha\beta}(r,r',\omega)$  and spectral functions,
 $\mathcal{P}_{\alpha\beta}(i,j,\omega) 
 = -\frac{1}{\pi}Im
 D_{\alpha\beta}(i,j,\omega)$ in real space for each 
 disorder configuration. This is then Fourier transformed to obtain the spectral functions
 in $(q,\omega)$ and then a disorder averaging is performed over relevant quantities.
\section{Local Single Particle Gap}
 In this section we provide the details of our method to obtain the local gap map, which
 is also used to show anti-correlation between one particle gap and two particle spectral
 weight in the system.
 The local single particle density of states (LDOS) for each site $i$, calculated from BdG 
 MF theory, is given by 
\begin{equation}
 N_{\omega}(r) = \frac{1}{N_s}\sum_{n} u_n^2(r)\delta(\omega-E_n)
 + v_n^2(r)\delta(\omega+E_n).
\end{equation}
 The local single particle gap $E(r)$ for each site $r$ is obtained from 
 $E(r) = \frac{\omega_p(r)-\omega_m(r)}{2}$, where $\omega_p(r)$ is the location of the 
 lowest energy peak in LDOS for $\omega>0$ and $\omega_m(r)$ is the location of the 
 highest energy peak in LDOS for $\omega<0$.
 In Fig. \ref{corr}(a) we have shown a sample LDOS for a particular site. The figure 
 also shows the location of $\omega_p(r)$ and $\omega_m(r)$ for this site and the 
 corresponding local gap $E(r)$ obtained from this LDOS.
\section{Covariance between Single Particle Gap and Two Particle Spectral Function}
To understand the spatial variation for the local two particle spectral function 
($P(i,i,\omega)$), we consider the integrated spectral weight of $P$, 
$F(i)=\int_0^{2E_{gap}} P(i,i,\omega) d\omega$.
We calculate the covariance between two experimentally observable quantities namely
two particle spectral function ($F$) and local single particle gap ($E$) as
\begin{equation}
 \text{cov}(F,E) = \langle FE\rangle - \langle F\rangle\langle E\rangle.
\end{equation}
In Fig. \ref{corr}(b) and (c) we show the covariance between $F$ and $E$ as a function of disorder, 
where $F$ has been calculated with different integration limits,
$F(i)=\int_{0.2\times2E_{gap}}^{2E_{gap}} P(i,i,\omega) d\omega$ and 
$F(i)=\int_{0.3\times2E_{gap}}^{2E_{gap}} P(i,i,\omega) d\omega $ respectively.
We find that with increasing the lower cut-off of the integration the anti-correlation 
between $F$ and $E$ at large disorder persists but it becomes weaker.

\end{document}